\documentclass[aps,prl,twocolumn,preprintnumbers]{revtex4}
\usepackage{amsmath,amssymb,bm,epsfig}

\begin{document}


\noindent
\textbf{Comment on ``Is There a ``Most Perfect Fluid'' Consistent 
with Quantum Field Theory?''}
\bigskip

Recently, Cohen~\cite{Cohen:2007qr} sketched a clever construction
which seems to invalidate the conjectured viscosity
bound \cite{Kovtun:2004de} while remaining within the domain of
well-defined relativistic quantum field theories.  The example
involves a metastable gas of heavy-light mesons in a gauge theory with
a large number of colors and flavors.  Here we note that in the
example of Ref.~\onlinecite{Cohen:2007qr}, the viscosity and the
entropy density exist in non-overlapping length-scale regimes.  Hence
the system does not constitute a counter-example to the viscosity
bound, if one makes a reasonable assumption that the viscosity and the
entropy density exist in the same physical regime.

In fact, the lifetime of this metastable system, as noted in
Ref.~\onlinecite{Cherman:2007fj}, may scale as a power of $\xi$ in the
limit $\xi\to\infty$, in unit of the time a meson travels the mean
interparticle separation.  It is due to the recombination of $O(1)$
particles into a bound cluster.  One the other hand, the number of
flavors scales as $\exp(\xi^4)$.  The entropy density is a sensible
concept only in a volume which contains all particle species, which
means that the linear size of the volume has to be larger than
$\exp(\frac13 \xi^4)$, in unit of the interparticle separation.  Any
measurement of the viscosity over such a long distance will take much
longer time than the lifetime of the system.

The example constructed in Ref.~\onlinecite{Cohen:2007qr} is a
remarkable system where the hydrodynamic regime is achieved at a
distance where the thermodynamic limit is still far from being
reached.  One seems unable to measure the viscosity over distance
scales associated with the thermodynamic limit.  It seems reasonable,
thus, to refrain from applying the viscosity bound to such a system.

It remains true that within nonrelativistic quantum mechanics (without
concerning about relativistic completion) one can evade the viscosity
bound by increasing the number of particle
species~\cite{Kovtun:2004de,Cohen:2007qr}.

I thank A.~Cherman, T.~D.~Cohen, P.~M.~Hohler, and G.~D.~Moore for
discussions.  This work is supported, in part, by DOE Grant No.\
DE-FG02-00ER41132.

\bigskip

\noindent
D.~T.~Son\\
$~~~$ Institute for Nuclear Theory\\
$~~~$ University of Washington\\
$~~~$ Seattle, Washington 98195-1550, USA\\


\begin{thebibliography}{9}

\bibitem{Cohen:2007qr}
  T.~D.~Cohen,
  Phys.\ Rev.\ Lett.\  {\bf 99}, 021602 (2007).

\bibitem{Kovtun:2004de}
  P.~Kovtun, D.~T.~Son and A.~O.~Starinets,
  Phys.\ Rev.\ Lett.\  {\bf 94}, 111601 (2005).

\bibitem{Cherman:2007fj}
  A.~Cherman, T.~D.~Cohen and P.~M.~Hohler,
  arXiv: 0708.4201 [hep-th].




\end{thebibliography}
\end{document}